\documentclass{aa}  

\usepackage{lscape}
\usepackage{graphicx}
\usepackage{multirow}
\usepackage{array}
\usepackage{txfonts}
\usepackage{todonotes}
\usepackage{footmisc}
\usepackage{hyperref}
\begin{document}

   \title{The solar cycle 25 multi-spacecraft solar energetic particle event catalog of the SERPENTINE project}

   \subtitle{}

   \author{N. Dresing\inst{1}
          \and
          A. Yli-Laurila \inst{1}
          \and
          S. Valkila \inst{1}
          \and    
          J. Gieseler \inst{1}
          \and
          D. E. Morosan \inst{1}
          \and
          G.~U. Farwa \inst{1}
          \and
          Y. Kartavykh \inst{3}
          \and
          C. Palmroos \inst{1}
          \and 
          I. Jebaraj \inst{1}
          \and
          S. Jensen \inst{3}
         \and
          P. Kühl \inst{3}
          \and
          B. Heber \inst{3}
          \and
          F. Espinosa \inst{4}
          \and
          R. G\'omez-Herrero \inst{4}
           \and
          E. Kilpua \inst{2}
         \and
          V.-V. Linho \inst{2}
          \and
          P. Oleynik \inst{1}
         \and 
          L.A. Hayes \inst{9}
          \and
          A. Warmuth \inst{5}
          \and
          F. Schuller \inst{5}
          \and
          H. Collier \inst{7, 8}
          \and 
          H. Xiao \inst{7}
           \and
          E. Asvestari \inst{2}
          \and
          D. Trotta \inst{6}
          \and
          J.G. Mitchell \inst{10}
          \and
          C.M.S. Cohen \inst{11}
        \and
          A.W. Labrador \inst{11}
          \and
          M.E. Hill \inst{12}
          \and
          R. Vainio \inst{1}
          }

   \institute{Department of Physics and Astronomy, University of Turku, Finland
              \email{nina.dresing@utu.fi}
         \and
         Department of Physics, University of Helsinki, P.O. Box 64, FI-00014 Helsinki, Finland  
         \and
         Institute of Experimental and Applied Physics, Kiel University, Kiel, Germany
         \and
         Universidad de Alcalá, Department of Physics, Space Research Group, Alcalá de Henares, Spain
         \and
         Leibniz-Institut für Astrophysik Potsdam (AIP), An der Sternwarte 16, D-14482 Potsdam, Germany
         \and
        The Blackett Laboratory, Department of Physics, Imperial College London, London, SW7 2AZ, UK
        \and
        Fachhochschule Nordwestschweiz, Bahnhofstrasse 6, 5210 Windisch, Switzerland
        \and
        ETH Z\"{u}rich,
        R\"{a}mistrasse 101, 8092 Z\"{u}rich Switzerland
        \and 
        European Space Agency, ESTEC,
        Keplerlaan 1 - 2201 AZ, Noordwijk, The Netherlands
        \and
        NASA Goddard Space Flight Center, Greenbelt MD 20771, USA
        \and
        California Institute of Technology, Pasadena, CA 91125, USA
        \and
        Johns Hopkins University Applied Physics Laboratory, Laurel, MD  20723, USA
        }

   \date{}

% \abstract{}{}{}{}{} 
% 5 {} token are mandatory
 
  \abstract
   {The {\it Solar energetic particle analysis platform for the inner heliosphere} (SERPENTINE) project, funded through the H2020-SPACE-2020 call of the European Union’s Horizon 2020 framework programme, employs measurements of the new inner heliospheric spacecraft fleet to address several outstanding questions of the origin of solar energetic particle (SEP) events. The data products of SERPENTINE include event catalogs, which are provided to the scientific community. }
   {In this paper, we present SERPENTINE's new multi-spacecraft SEP event catalog for events observed in solar cycle 25. Observations from five different viewpoints are utilized, provided by Solar Orbiter, Parker Solar Probe, STEREO~A, BepiColombo, and the near-Earth spacecraft Wind and SOHO. The catalog contains key SEP parameters for $25-40$~MeV protons, $\sim1$~MeV electrons, and $\sim100$~keV electrons. Furthermore, basic parameters of the associated flare and type-II radio burst are listed, as well as the coordinates of the observer and solar source locations.}
   {An event is included in the catalog if at least two spacecraft detect a significant proton event with energies of $25-40$~MeV. SEP onset times are determined using the Poisson-CUSUM method. SEP peak times and intensities refer to the global intensity maximum. If different viewing directions are available, we use the one with the earliest onset for the onset determination and the one with the highest peak intensity for the peak identification. We furthermore aim at using a high time resolution to provide most accurate event times. Therefore, time averaging of the SEP intensity data is only applied if necessary to determine clean event onsets and peaks. Associated flares are identified using observations from near Earth and Solar Orbiter. Associated type II radio bursts are determined from ground-based observations in the metric frequency range and from spacecraft observations in the decametric range.}
   {The current version of the catalog contains 45 multi-spacecraft events observed in the period from Nov 2020 until May 2023, of which 13 were widespread events and four were classified as narrow-spread events. Using X-ray observations by GOES/XRS and Solar Orbiter/STIX, we were able to identify the associated flare in all but four events. Using ground-based and space-borne radio observations, we found an associated type-II radio burst for 40 events. In total, the catalog contains 142 single event observations, of which 20 (45) have been observed at radial distances below 0.6~AU (0.8~AU). It is anticipated to extend the catalog in the future.}
   {}


   \maketitle
%
%-------------------------------------------------------------------

%%%%%%%%%%%%%%%%%%%%%%%%%%%%%%%%%%%%%%%%%%%%%%%%%%%%%%%%%%%%%%%%%%%%%%%%
\section{Introduction} \label{sec:intro}
Solar energetic particle (SEP) events are large outbursts of energetic particle radiation from the Sun associated with solar eruptions, that is, solar flares and coronal mass ejections (CMEs). They can be classified based on their key observational properties, for example, in impulsive and gradual classes \citep[e.g.,][]{Reames1999}. The classical picture relates impulsive events with flares and gradual events with fast and wide CMEs. While the broad lines of the origin of SEP events are mostly understood, many of the key aspects of the acceleration and transport of ions and electrons in these events have remained elusive. The main obstacle preventing us from achieving a complete understanding has been the lack of observational coverage in the heliosphere: solar eruptions often fill a major part of the inner heliosphere with energetic particles, and using a scarce set of observing spacecraft does not allow one to conclusively separate the effects of particle transport from the properties of the source.

With the launch of the new space missions Solar Orbiter \citep{Muller2020} and Parker Solar Probe \citep[Parker;][]{Fox2016}, a unique era for multi-spacecraft observations of SEP events has begun. Combined with established space missions near Earth, such as the SOlar and Heliospheric Observatory \citep[SOHO;][]{Domingo1995} and Wind \citep{Ogilvie1997}, the ahead spacecraft of the Solar Terrestrial Relations Observatory \citep[STEREO~A;][]{Kaiser2008} as well as the BepiColombo mission on its cruise to Mercury \citep{Benkhoff2021}, these spacecraft form an unprecedented fleet during the rising phase of Solar Cycle 25. The various orbits of the missions provide ever-changing spacecraft constellations covering different heliocentric distances and heliolongitudes. Some SEP events are observed over wide longitudinal ranges in the inner heliosphere \citep[e.g.,][]{Kollhoff2021, Dresing2023}, others by a fleet of closely spaced observers, allowing the study of local effects \citep[e.g.,][]{Lario2022, Palmerio2024}. Sometimes different spacecraft are radially aligned, permitting to study radial dependencies, especially during interplanetary CME (ICME) and interplanetary (IP) shock crossings \cite[e.g.,][]{Kilpua2021, Trotta2023,Trotta2024}. Other events are observed by magnetically aligned spacecraft, that is when they are situated nearly on the same Parker spiral magnetic field connecting them to the same region at the Sun \citep[e.g.,][]{Rodriguez-Garcia2023b, Kollhoff2023}. Such constellations are ideal to investigate radial effects of parallel transport of SEPs along the same magnetic field line. Inner heliospheric observations can be complemented by SEP measurements at farther distances, such as Mars \citep[e.g.,][]{Palmerio2022, Guo2023}, e.g., detection on the surface \citep[][]{Hassler2012} and in orbit by the MAVEN spacecraft \citep[][]{Jakosky2015}. 

This powerful spacecraft fleet opens new opportunities to answer many outstanding questions on the generation of SEP events. 
What is the reason for widespread SEP events that show SEP distributions up to all around the Sun as recorded by spacecraft located at various longitudes? How are SEPs of different species and energies accelerated? What are the relative roles of the associated flare and CME-driven shock in particle acceleration and spatial SEP distribution? What is the role of particle transport in spreading the particles over wide longitudinal regions?

The new knowledge generated with these unprecedented observations will not only further our understanding of the fundamental processes, dominating several astrophysical systems, but eventually lead to better forecasting abilities of space-weather-relevant SEP events at Earth or elsewhere in the solar system.  

This paper describes the building and content of a multi-spacecraft SEP event catalog of SEP events detected in solar cycle 25 by inner heliospheric space missions, that is, Solar Orbiter, Parker Solar Probe, STEREO~A, BepiColombo, and the near-Earth spacecraft Wind and SOHO. 
The presented catalog was compiled within the scope of the {\it Solar energetic particle analysis platform for the inner heliosphere} (SERPENTINE\footnote{\url{https://serpentine-h2020.eu}}) project, which is funded through the European Union's Horizon 2020 framework programme under the H2020-SPACE-2020 call topic addressing \emph{Scientific data exploitation}.

The main objective of SERPENTINE is to uncover the primary causes of large, gradual, and widespread SEP events, that is, the respective role of the broad sources and efficient cross-field particle transport in their genesis. The project tackles the challenge through comprehensive analyses, both case studies and statistical investigations, of historical and current SEP measurements and solar context observations \citep{Kollhoff2021, Dresing2022, Dresing2023, Rodriguez-Garcia2023a, Rodriguez-Garcia2023b, Wijsen23, Jebaraj2023, Jebaraj2023b, Trotta2022a, Dimmock2023, Lorfing2023, Kilpua2023, Trotta2023, Trotta2023b, Trotta2023c, Trotta2024}. One of the most important outcomes of the project is the public release of data analysis tools \citep{Kouloumvakos2022b, Palmroos2022, Price2022, Trotta2022, Gieseler2023, Kouloumvakos2023} and catalogs of SEP events, interplanetary shocks, and CMEs for historical and solar cycle 25 events. The tools and catalogs are built for easy access and are released in the hope of broad use by the heliophysics community. Altogether, five catalogs, two based on Helios data and three based on modern observations, have been released through the project data server (\url{https://data.serpentine-h2020.eu}). Additionally, the presented SEP catalog is accessible and citable as \citet{Dresing2024zenodo} through Zenodo\footnote{\url{https://doi.org/10.5281/zenodo.10732268}}.

This paper presents the SERPENTINE catalog of the SEP events of solar cycle 25 to date. The catalog focuses on energetic SEP events, producing at least 25 MeV protons, potentially relevant for space weather, and aims to provide a comprehensive resource to exploit the novel multi-spacecraft observations of solar cycle 25.

The paper is organized as follows. In Section \ref{sec:data}, we introduce the data and energetic particle instruments used; in Section \ref{sec:selection_criteria}, we explain the selection criteria for listed events; and in Section \ref{sec:catalog} we present the contents of the catalog and how the parameters were determined. Section \ref{sec:results} presents the results, and Section \ref{sec:summary} summarizes the work.

%%%%%%%%%%%%%%%%%%%%%%%%%%%%%%%%%%%%%%%%%%%%%%%%%%%%%%%%%%%%%%%%%%%%%%%%
\section{Data and instrumentation} \label{sec:data}
We utilize energetic particle measurements taken by six space missions: Solar Orbiter, Parker, STEREO~A, Wind, SOHO, and BepiColombo. 
In order to cover the three different SEP energy / species combinations of the catalog, $>$25-MeV protons as well as $\sim$100-keV and $\sim1$-MeV electrons, we use the following instruments: the Energetic Particle Detector \citep[EPD;][]{Rodriguez-Pacheco2020} instrument suite of Solar Orbiter, the Integrated Science Investigation of the Sun \citep[IS$\odot$IS;][]{McComas2016} suite of Parker, the Solar Electron Proton Telescope \citep[SEPT;][]{Muller-Mellin2008} and the High Energy Telescope \citep[HET,][]{vonRosenvinge2008} of the STEREO mission, the Three-Dimensional Plasma and Energetic Particle Investigation \citep[3DP;][]{Lin1995} of the Wind spacecraft, the Energetic and Relativistic Nuclei and Electron \citep[ERNE;][]{Torsti1995} experiment and the Electron Proton Helium Instrument (EPHIN), which is part of the Comprehensive Suprathermal and Energetic Particle Analyser \citep[COSTEP;][]{Muller-Mellin1995} suite of SOHO, as well as the Solar Intensity X-Ray and Particle Spectrometer \citep[SIXS;][]{Huovelin2020} on board BepiColombo's Mercury Planetary Orbiter (MPO).

For each of the three energy / species combinations, we aim at providing the SEP parameters at the same energy range as observed by the different spacecraft (S/C). Because of the different instrumentation on board the various S/C, no perfect overlap is possible. Table~\ref{tab:instruments} summarizes the chosen S/C and instruments (column 1), the energy channels and corresponding mean energies $\langle E\rangle$ (columns 3-5) that were used in the catalog, and the available viewing directions of the instruments (column 2). We note that the data product of Parker/EPI-Lo has changed in June 2021, which leads to a changed energy range and different energy channels to be used. In the case of BepiColombo/SIXS, only effective energies of the channels can be determined because of the complex instrument response functions. The energy-range limits provided for BepiColombo/SIXS in columns (2) and (4) of Table \ref{tab:instruments} therefore represent the effective mean energies of the used energy channels. This means that the actual energy channels are significantly wider. For SOHO/EPHIN, it should be noted that the instrument was switched to failure modes D and E in 2017, which excluded the two deepest detectors from any coincidence logic. This resulted in a significant widening of the electron channel E1300. Therefore, the effective mean energy has a strong spectral dependence. The corresponding value of 1.47~MeV shown in Table~\ref{tab:instruments} has been calculated for a spectral index of -2\footnote{Assuming a spectral index of -3 results in a mean energy of 1.02~MeV, while a spectral index of -1.5 yields a mean energy of 1.95~MeV.}. This corresponds to the average slope of the events considered in this paper (see Sect. \ref{sec:ratios}).

Near-relativistic electron measurements, like the $\sim$100-keV electron channel, employed in our catalog, are typically realized through a single-detector measurement and the magnet-foil principle, where a foil is used to stop protons and ions from reaching the detector. However, ions that penetrate the foil can contaminate the electron measurements, especially during periods when ion fluxes are large compared to the electron fluxes \cite[e.g.,][]{Wraase2018}. This is more often the case during SEP peak phases than during SEP onsets due to the usually earlier arrival times of electrons. Therefore, we checked all $\sim$100-keV electron peaks by comparing the electron-intensity time series with those of ions in the contaminating energy ranges and excluded the peaks from the catalog if in doubt. 
\\
\begin{table*}
\caption{\label{tab:instruments} Energetic particle instruments, their available viewing directions, and energy channels used for the catalog.}
\centering
%\todo[inline]{BH: If EPHIN or Erne are looking sunward depends on the orientation of the spacecraft. Thus it could view also perpendicular to the nominal field line.
%The same is true for HET and SEPT on STEREO. Stefan: can you look up the mean energy of EPHIN E1300.}
\begin{tabular}{lllll}
\hline\\
(1) & (2) & (3) & (4) & (5) \\[0.5mm]
Spacecraft / Instrument& No. of available viewing directions &  $>$ 25 MeV p &  $\sim1$~MeV e &  $\sim$ 100 keV e\\[1mm]
\hline \\
SOHO / ERNE / HED &1 \tablefootmark{a} & CH 3--4:  &  & \\
 &(sunward)& 25 -- 40 MeV & & \\
    &  &   $\langle E\rangle=31.6$~MeV &  & \\
SOHO / EPHIN & 1 \tablefootmark{a}& & E1300:  & \\
 &(sunward)& & 0.67 -- 10.4 MeV & \\
   &  & &  $\langle E\rangle=1.47$~MeV &   \\
Wind / 3DP & 8 &  & & CH 3:  \\
 &sectors 0--7& & & 82.26 -- 135 keV \\
      & & & & $\langle E\rangle=105.4$~keV \\
STEREO / HET & 1 \tablefootmark{b} & CH 5--8:  & CH 0--1:  & \\
 &(sunward)& 26.3 -- 40.5 MeV & 0.7-2.8 MeV & \\
   &  &  $\langle E\rangle=32.6$~MeV &  $\langle E\rangle=1.96$~MeV & \\
STEREO / SEPT &4& & & CH 6--7:  \\
 &(sun\tablefootmark{b}, anti-sun\tablefootmark{b}, north, south)& & & 85 -- 125 keV \\
     & & & & $\langle E\rangle=103.1$~keV \\
SolO / EPD / HET & 4 & CH 19--24:  & CH 0-1:  & \\
 &(sun\tablefootmark{c}, anti-sun\tablefootmark{c}, north, south)& 25.09 -- 41.18 MeV & 0.45--2.4 MeV & \\
  &  &  $\langle E\rangle=32.1$~MeV &  $\langle E\rangle=1.1$~MeV & \\
SolO / EPD / EPT & 4 & & & CH 14--18:  \\
 &(sun\tablefootmark{c}, anti-sun\tablefootmark{c}, north, south)& & & 85.6 -- 130.5 keV \\
    & & & & $\langle E\rangle=105.7$~keV \\
Parker / EPI-Hi / HET &2& CH 8--9: & CH 3--4: & \\
 &(A: sunward\tablefootmark{c}, B: anti-sunward\tablefootmark{c})& 26.91 -- 38.05 MeV & 0.71 -- 1.41 MeV & \\
 & &  $\langle E\rangle=32.0$~MeV & $\langle E\rangle=1.0$~MeV  &    \\
Parker / EPI-Lo & 8 wedges\tablefootmark{d} (w0-w7)& & & CH F 4--5: \\
 &w3: sunward\tablefootmark{c}; w7: anti-sunward\tablefootmark{c} & & & 65--153 keV\tablefootmark{e} \\
    & & & & $\langle E\rangle=98.9$~keV \\
 BepiColombo / SIXS & 3(4)\tablefootmark{f}& CH 8--9: & CH 5--6: & CH 2: \\
  &side 0--3\tablefootmark{f}& 25.1 -- 37.3 MeV\tablefootmark{g} & 0.96 -- 2.24 MeV\tablefootmark{g} & 106 keV\tablefootmark{g} \\
   &  & $\langle E\rangle=35$~MeV & $\langle E\rangle=1.5$~MeV & \\
\hline
\end{tabular}
\tablefoot{\\
\tablefoottext{a}{Along the nominal Parker spiral. Since 2003 the viewing direction changes four times a year (\url{https://soho.nascom.nasa.gov/data/ancillary/attitude/roll/nominal_roll_attitude.dat})}
\tablefoottext{b}{Along the nominal Parker spiral. Since mid-2015 (after the superior conjunction), perpendicular to it.}
\tablefoottext{c}{Along the nominal Parker spiral.}
\tablefoottext{d}{Only wedges 3 and 7 were used.}
%\tablefoottext{e}{Only counts available.}
\tablefoottext{e}{Data product changed. Before 14 June 2021, CH F 3--4: 84--138 keV was used.}
\tablefoottext{f}{Side 4 is not used because it is blocked by the sunshield during the cruise phase. Side 3 is often excluded due to noise issues.}
\tablefoottext{g}{Only effective energies available.}
}
\end{table*}

Key parameters of the associated flare rely on GOES X-ray observations as provided by SolarMonitor\footnote{\url{https://solarmonitor.org}}.
The presence of metric type II radio bursts is also investigated using event lists provided by the Space Weather Prediction Center (SWPC) \footnote{\url{https://www.swpc.noaa.gov/products/solar-and-geophysical-event-reports}}. SWPC collects reports from contributing stations from the ground that provide 24-hour coverage. These stations are Culgoora spectrograph, Learmonth spectrograph, Holloman Solar Observatory, and San Vito Solar Observatory, and they form the Radio Solar Telescope Network (RSTN).

%%%%%%%%%%%%%%%%%%%%%%%%%%%%%%%%%%%%%%%%%%%%%%%%%%%%%%%%%%%%%%%%%%%%%%%%
\section{Selection criteria} \label{sec:selection_criteria}
\begin{figure*}[t]  
        \includegraphics[width=0.6\textwidth]{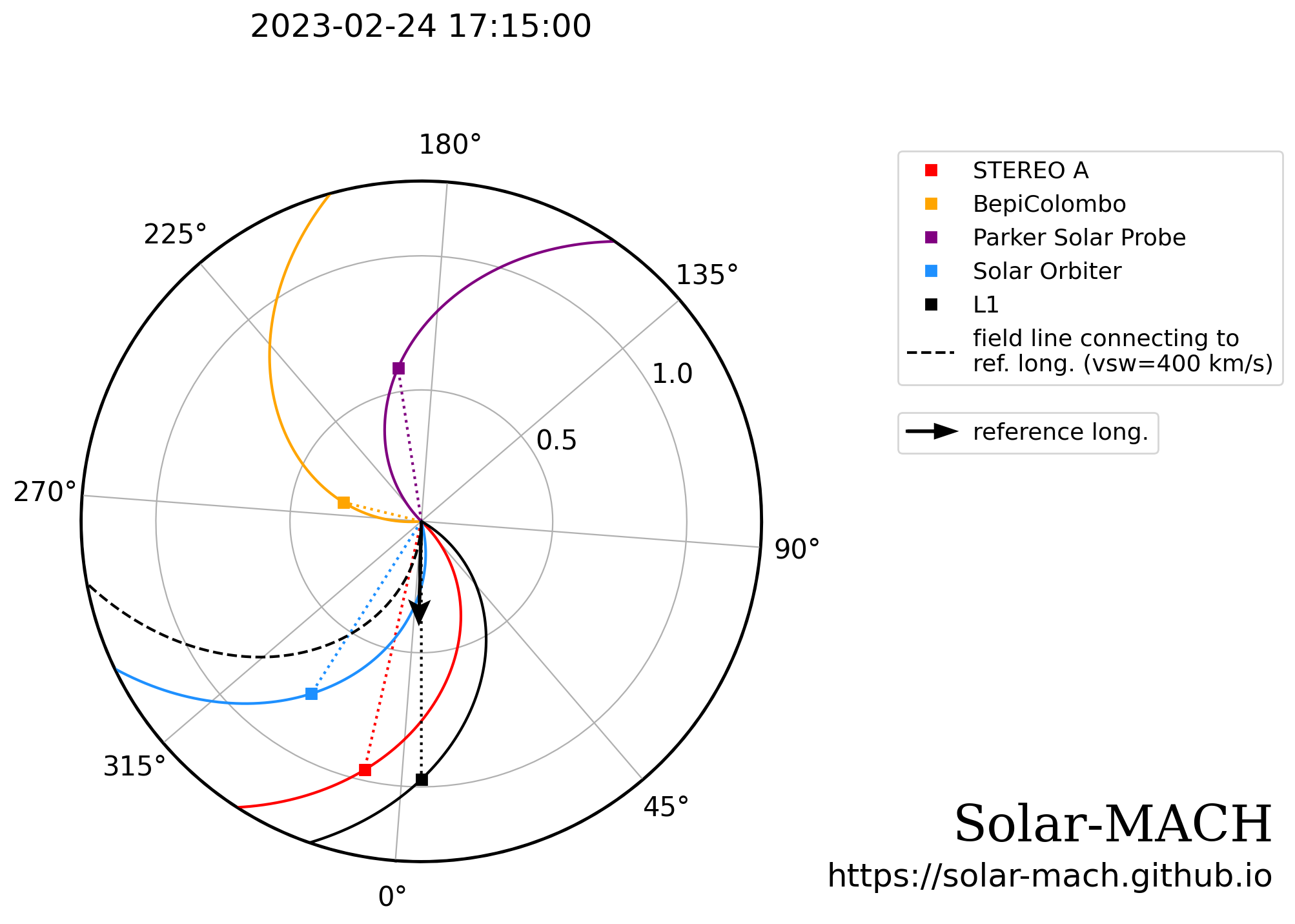} \\ % solar mach plot
        \centering
        \includegraphics[width=0.9\textwidth]{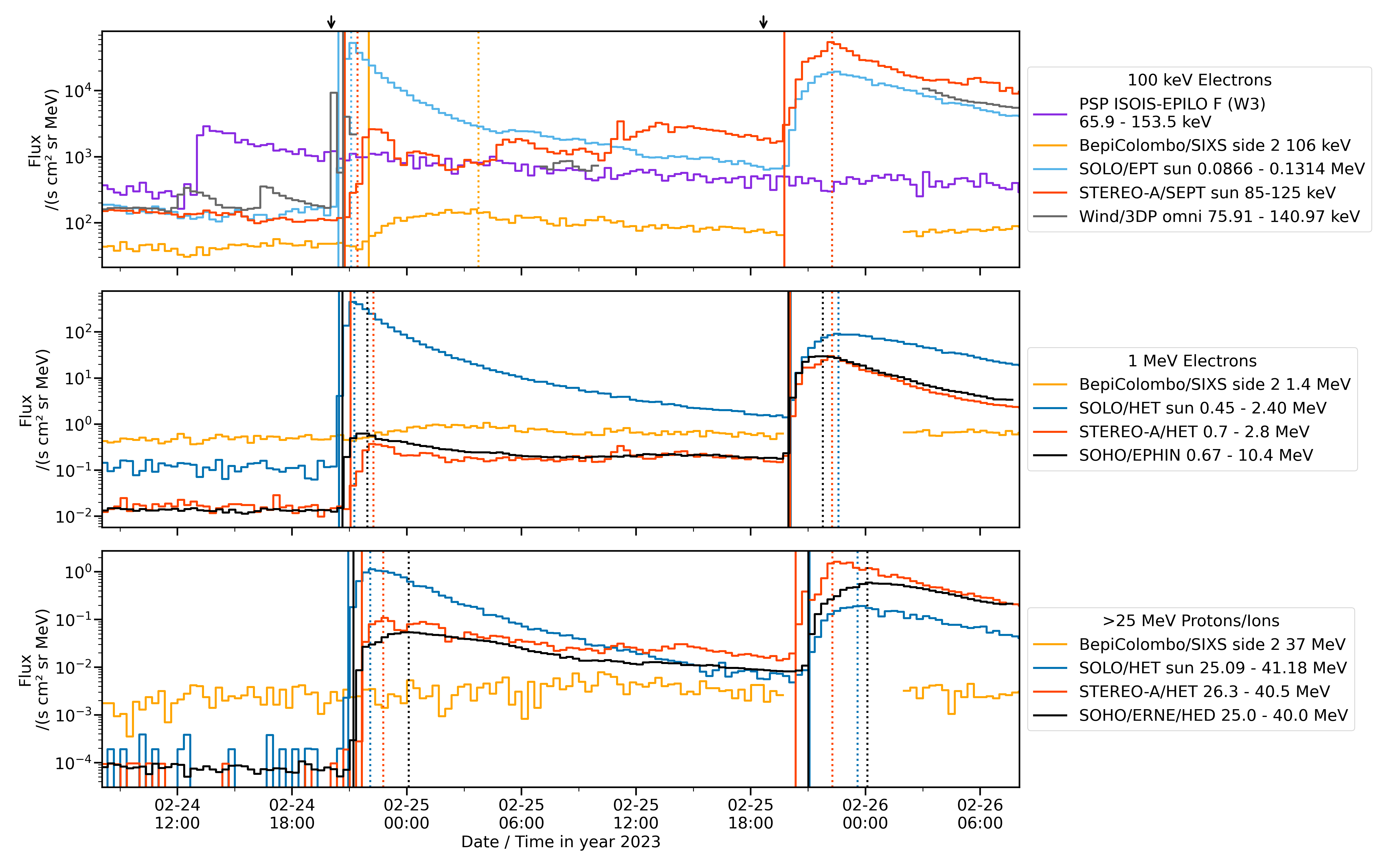}  % multi-sc plot
         
        \caption{Multi-spacecraft events of 24 and 25 Feb 2023. The top plot presents the longitudinal configuration of spacecraft in the heliographic equatorial plane with respect to the longitude of the associated flare (black arrow) for the event starting on 24 Feb. The plot has been produced with the Solar-MACH tool \citep{Gieseler2023}. The colored spirals denote the magnetic field lines connecting each observer with the Sun, the black dashed spiral represents the field line connecting with the flare. The bottom plot shows the intensity time series of $\sim100$-keV electrons (top), $\sim1$-MeV electrons (middle), and $>25$-MeV protons (bottom). Parker did not observe the events but a previous event happening earlier on 24 Feb. Since there is no obvious electron foreground in the EPI-Lo electron measurements (top panel) during the second event, the counts during this time period are background, assumed to be primarily caused by galactic cosmic rays. Parker/HET data were not available during the shown time period. The black arrow on top of the upper panel denotes the times of the associated flares. Solid (dotted) lines mark the onset (peak) times with colors corresponding to those of the spacecraft time series.} 
        \label{fig:multi-sc1}
\end{figure*}

\begin{figure*}[]  
        \includegraphics[width=0.6\textwidth]{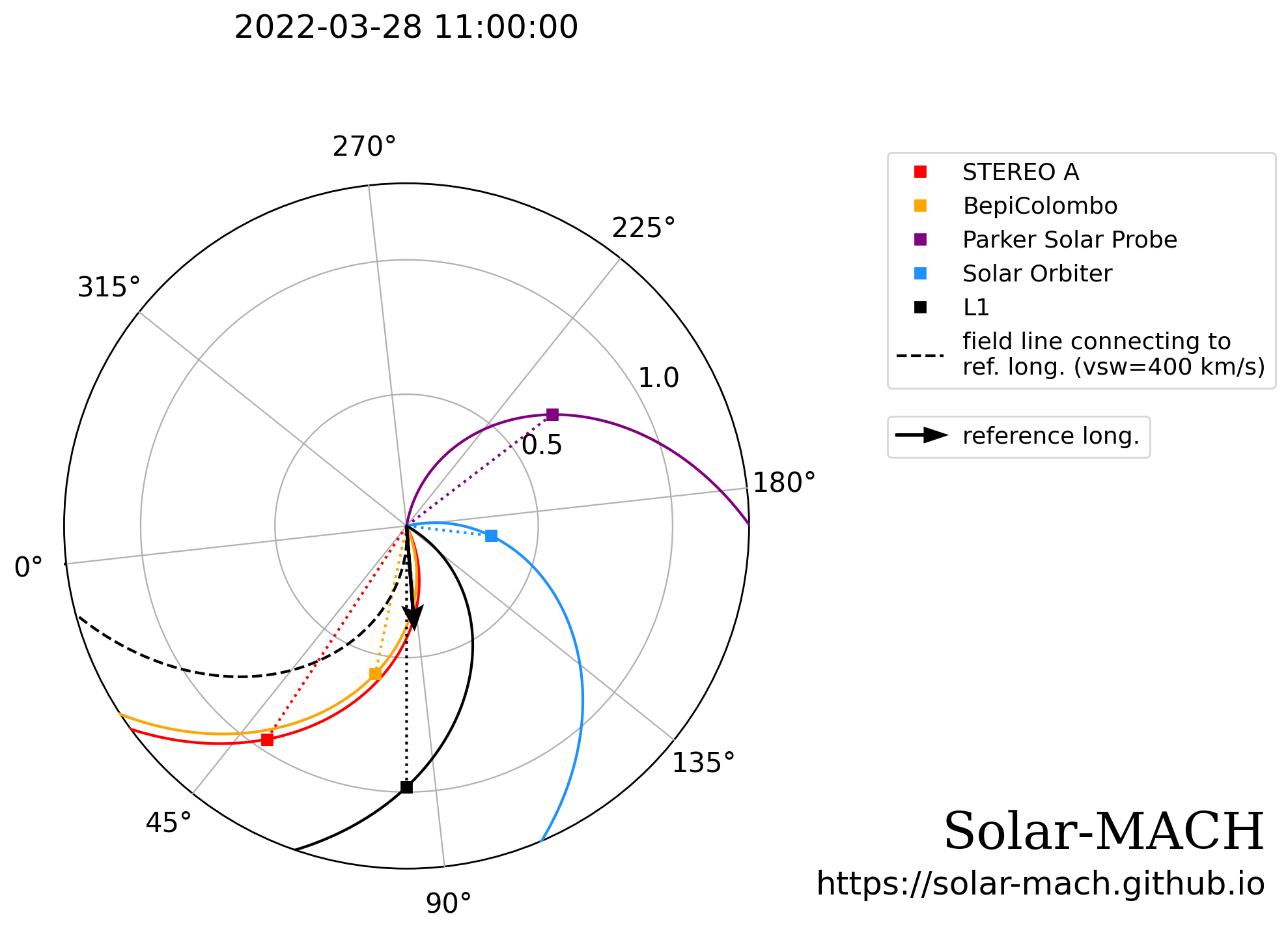} \\ % solar mach plot
        \centering
        \includegraphics[width=0.9\textwidth]{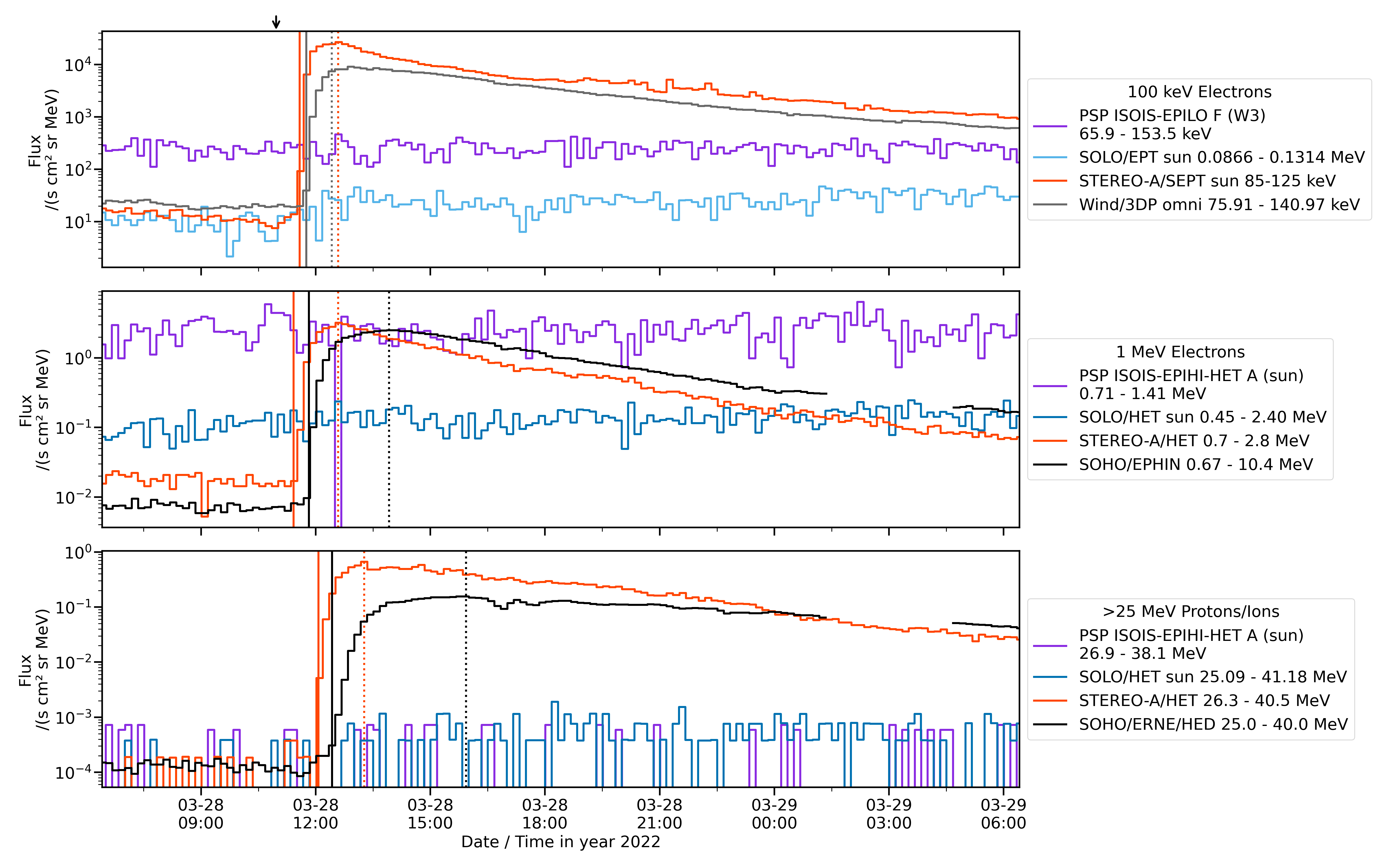}  % multi-sc plot
         
        \caption{Figure structure as in Fig.~\ref{fig:multi-sc1}, but for the event of 28 Mar 2022. This event was only observed by STEREO~A and near-Earth spacecraft (Wind and SOHO), but not at Solar Orbiter nor Parker (BepiColombo/SIXS data were not available). The science case of this event is 'narrow-spread event'.} 
        \label{fig:multi-sc2}
\end{figure*}

Following previous SEP event catalogs \cite[e.g.,][]{Richardson_Cane2010, Richardson2014}, we base our event selection on energetic proton events observed in the energy range of $\sim~25-40$ MeV (see Table~\ref{tab:instruments}, column 3). An event is identified based on visual inspection of 1-hour averaged data, requiring a significant increase above background.
Aiming at a multi-spacecraft catalog, we then select an event if it was observed by at least two different spacecraft. Due to the constantly changing spacecraft constellations, this includes both, events observed over wide longitudinal regions (widespread events, cf.\ Figure~\ref{fig:multi-sc1}) and those observed only in a narrow longitudinal sector (cf. Figure~\ref{fig:multi-sc2}). The latter also includes spacecraft constellations with radial alignments or those where different spacecraft are magnetically aligned, that is, they are situated along the same nominal Parker spiral connecting to the Sun (cf. BepiColombo and STEREO~A in Figure~\ref{fig:multi-sc2}). Therefore, one column of the catalog provides cases of potential scientific interest based on the multi-spacecraft constellation of the observing spacecraft. 

The multi-spacecraft SEP events are identified based on the observations of temporal coincident enhancements of energetic particle fluxes at the different spacecraft. Key information about the associated flare is also listed when a flare was reported in the Earth's or Solar Orbiter's visible hemisphere. 

\section{Contents of the catalog and parameter determination} \label{sec:catalog}
\begin{figure*}[h!]  
\centering
        \includegraphics[width=0.8\textwidth]{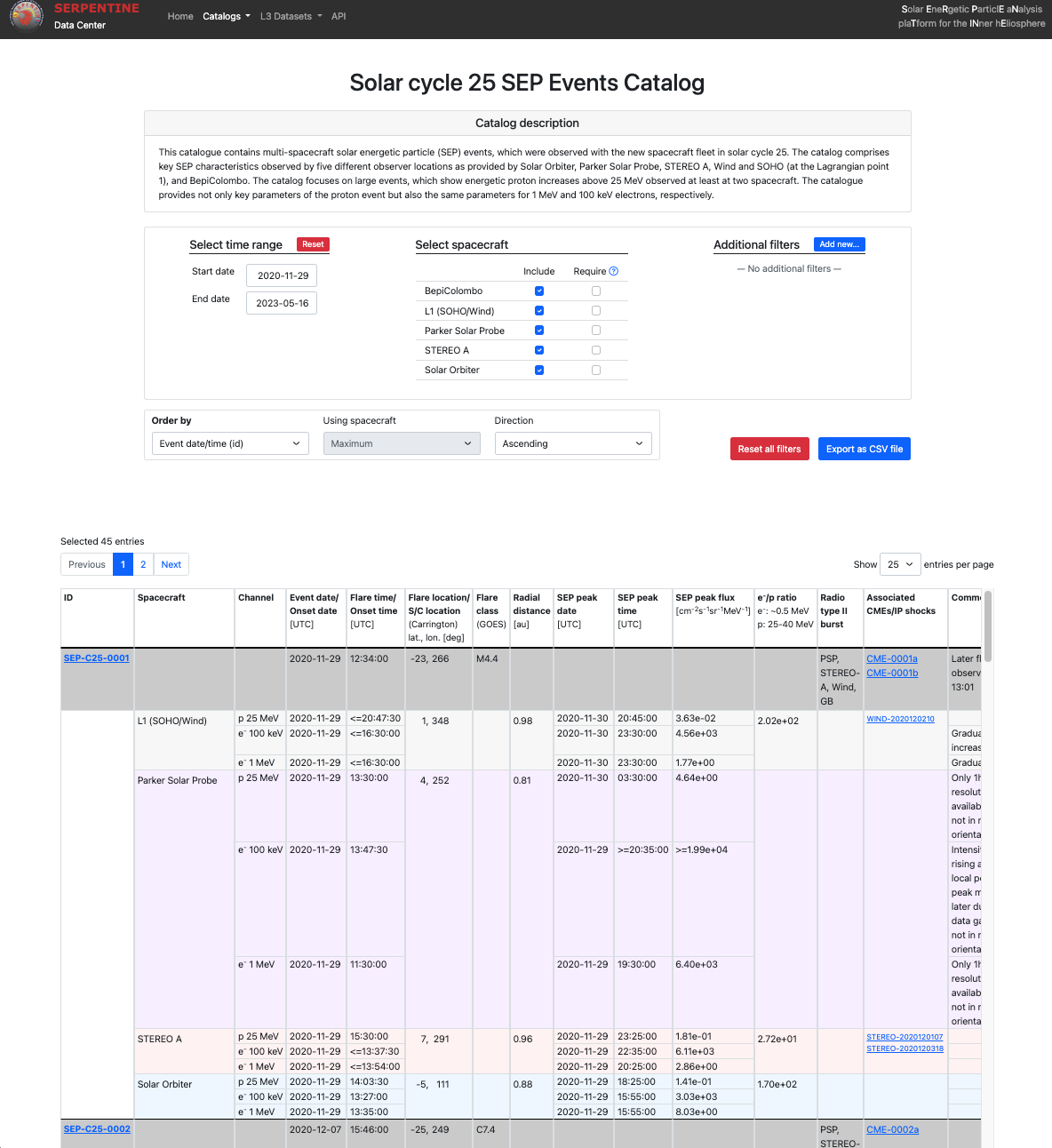} 
        \caption{The starting page of the multi-spacecraft SEP event catalog at the SERPENTINE datacenter. Various filtering options can be applied by the user. Below, the resulting list of events including several of the main event parameters are shown.} 
        \label{fig:data_center}
\end{figure*}

\begin{figure*}[]  
\centering
        \includegraphics[width=0.8\textwidth]{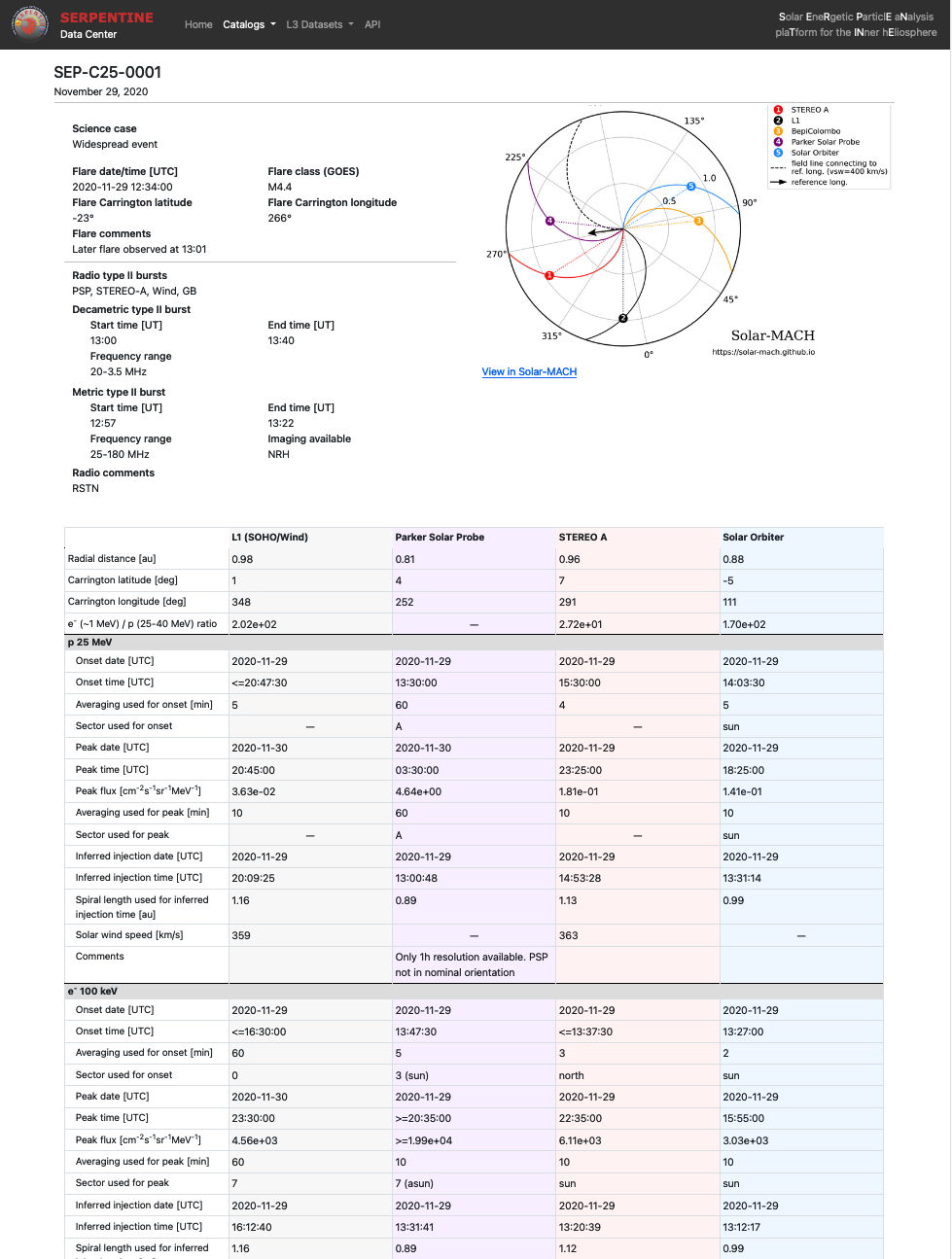} 
         
        \caption{Detailed view of a selected event at the SERPENTINE datacenter. This page shows all parameters of the selected event.} 
        \label{fig:data_center_event}
\end{figure*}

%%%%%%%%%%%%%%%%%%%%%%%%%%%%%%%%%%%%%%%%%%%%%%%%%%
The current version of the catalog is archived and citable as \citet{Dresing2024zenodo} through Zenodo\footnote{\url{https://doi.org/10.5281/zenodo.10732268}}, where newer versions will be added after every update. 
The latest version of the catalog can also be accessed via the SERPENTINE data center\footnote{\url{https://data.serpentine-h2020.eu/catalogs/sep-sc25/}} along with (and linked to) further data products and catalogs provided by the SERPENTINE project. In addition to downloading the catalog from the data center, the platform provides various filtering options. Figure~\ref{fig:data_center} shows a screenshot of the catalog's starting page, which contains only a selection of parameters for each event. After clicking an event ID, a detailed event page opens (see Fig.~\ref{fig:data_center_event}) that presents all event parameters as well as a plot showing the S/C constellation and their nominal magnetic connections with the Sun as determined by the Solar-MACH tool \citep{Gieseler2023}. 
For those events where an associated flare was identified (see details below), its position is added as a reference (arrow) to the Solar-MACH plot. We note that the Solar-MACH tool provides further coordinate information such as the locations of the S/C magnetic footpoints estimated via ballistic backmapping (using the measured solar wind speed or assuming a solar wind speed of 400~km/s) as well as the longitudinal separation angles with the flare location. Those values can be retrieved from the Solar-MACH webpage\footnote{\url{https://solar-mach.streamlit.app}}, when clicking the link below the plot on the catalog page. We note, that the SERPENTINE project provides further Python-based tools on the SERPENTINE Hub\footnote{\url{https://hub-serpentine.rahtiapp.fi/hub/}}, which can be used to perform more in-depth analyses of SEP events. These include tools to load and visualize various SEP measurements as well as a magnetic connectivity tool combining ballistic backmapping with magnetic field extrapolations in the lower corona based on a Potential Field Source Surface (PFSS) model \citep[see also, ][]{Gieseler2023, Palmroos2022}. 

For each SEP event, the catalog provides various SEP parameters as obtained by the five different observer locations as well as flares based on GOES and Solar Orbiter observations, and the presence of type II radio bursts along with their occurrence time and frequency range. The columns of the catalog for each SEP event are as follows:
\begin{enumerate}
%\item Event number: number of the event in the catalog, starting with 1, monotonically increasing.
\item Event ID: Unique identifier of each SEP event used for linking different SERPENTINE catalogs.
\item Science case: For example: 'widespread event'\footnote{A widespread event is defined as an event observed by at least one observer with a longitudinal separation $\Phi$ between observer's magnetic footpoint and flare longitude of $\geq$80 degrees, following \citet{Dresing2014}.}, 'narrow-spread event'\footnote{We define a narrow-spread event as an event that is not observed by an observer who has a longitudinal separation angle $\Phi\leq$80 degrees.}, 'magnetic alignment of STEREO~A and L1’, 'radial alignment of BepiColombo and L1'.
%\item Observer identifier: unique identifier for a specific S/C of the specific event, used for linking with other SERPENTINE catalogs.
%\item Observer: For each event the catalog contains one row for each of our five observer locations, that is, Solar Orbiter, Parker, STEREO~A, BepiColombo, and L1. Observations at L1 are a combination of Wind and SOHO measurements. 
\item Event start date: Date of the associated flare or earliest SEP onset (in case no flare observations are available).
%\item Flare date
\item Flare time\footnote{All times in the catalog (except the inferred SEP injection times) refer to their measurement at the spacecraft.} (UT)
\item Flare latitude\footnote{\label{coord_foot}All coordinates of the catalog are provided in the Carrington coordinate system} (°)
\item Flare longitude\footref{coord_foot}  (°)
\item Flare class (GOES)
\item Flare comments
\item Radio type II burst: Observer names\footnote{We use spacecraft names or 'GB' for ground-based observations of metric type II bursts.} are provided in case of a type II observation, otherwise left empty.
\item Decametric radio type II burst start time (UT)
\item Decametric radio type II burst stop time (UT)
\item Frequency range of decametric type II burst (MHz)
\item Metric radio type II burst start time (UT)
\item Metric radio type II burst stop time (UT)
\item Frequency range of metric type II burst (MHz)
\item Metric radio imaging available (yes/no)
\item Radio comments\footnote{Radio comments provide the instrument used for metric type II burst identification or spacecraft data availability notes.} 
\item ID of associated CMEs\footnote{These CMEs are listed in the SERPENTINE CME catalog at \url{https://data.serpentine-h2020.eu/catalogs/cme/}.}
\item Solar-MACH link: Link to a spacecraft constellation plot of the event at \url{https://solar-mach.github.io}. \\
\\
The following columns are provided for each spacecraft:\\
\item S/C latitude\footref{coord_foot} (°)
\item S/C longitude\footref{coord_foot} (°)
\item S/C radial distance from the Sun (au)
\item ID of associated interplanetary shocks\footnote{These shocks are searched in association with the events and are listed in the SERPENTINE IP shock catalog at \url{https://data.serpentine-h2020.eu/catalogs/shock-sc25/}.}\\
\\
%\item Solar-MACH link: A link to an S/C constellation plot for the time of the event using Solar-MACH{\footnote{\url{https://solar-mach.github.io}}} \citep{Gieseler2023}. If an associated flare is listed, its longitudinal position is included as a reference longitude in the plot.
The following columns are provided for each of the three particle species / energy combinations (i) $>25$-MeV protons, (ii) $\sim100$-keV electrons, and (iii) $\sim1$-MeV electrons:\\
\begin{enumerate}
\item Onset date: Date of the energetic particle onset at the S/C
\item Onset time: Time of the energetic particle onset at the S/C (UT)
\item Time averaging used for the onset determination (min)
\item Sector used for onset determination: This applies only to those instruments that provide directional energetic particle measurements, see Table~\ref{tab:instruments}
\item Peak date: Date of maximum intensity (UT)
\item Peak time: Time of maximum intensity (UT)
\item Peak flux: Maximum intensity ($\mathrm{(s~cm^2~sr~MeV)^{-1}}$) of the event in the selected energy channel
\item Averaging used for peak determination (min)
\item Sector used for peak determination: Same as (d) (see above)
\item Inferred injection date (UT) 
\item Inferred injection time (UT)\footnote{The uncertainty of the injection time is at least as high as the time averaging used to determine the corresponding onset time.}, based on the onset time of the selected species and energy channel considering the distance of the S/C and assuming a
nominal Parker spiral (see more detail in Sect.~\ref{sec:sep_parameters})
\item Path length (au) used for inferred injection time: Length of the nominal Parker spiral considering the distance of the S/C and using the measured solar wind speed
\item Solar wind speed (km/s) used to determine path length\footnote{Based on measurements at the S/C. If empty, a nominal value of 400~km/s was used.}
\item Event comments\footnote{In case of Parker the comments also contain a remark if the spacecraft was not nominally orientated.}
\end{enumerate}
\item Electron / proton ratio
\end{enumerate}

\subsection{Flare identification}\label{sec:flare}
Several sources were employed to find the most appropriate solar flare, which could produce the observed multi-spacecraft event. The majority of associated flares were found with the help of SolarMonitor, based on observations of the X-ray sensor (XRS) on board the GOES spacecraft covering the visible hemisphere of Earth. We also used other resources if no flare candidate was listed at SolarMonitor, that is the catalog of flares observed by the Hinode satellite \citep{Watanabe2012} \footnote{\url{https://hinode.isee.nagoya-u.ac.jp/flare_catalogue/}} and the Space Weather Prediction Center \footnote{\url{ftp://ftp.swpc.noaa.gov/pub/indices/events/}}. If the flare was not visible from Earth's viewpoint we used flare observations by Solar Orbiter's X-ray telescope STIX \citep{Krucker2020} as provided in the Solar Orbiter flare catalog\footnote{\url{https://github.com/hayesla/stix_flarelist_science}. The STIX flare list was created by the STIX ground software and data processing team \citep{stixdatacenter}}. 
Potentially corresponding flares were identified based on timing coincidence. In a first step, we looked for flares within a time window several hours before the first particle onset until this first onset. Afterward, the flare candidate was corroborated by its location with respect to the spacecraft constellation, assuming that the magnetically best-connected would observe the highest SEP intensities and earliest onset times. If there were several flares from the point of view of their location, which could serve as a potential source, the preference was given to the most powerful one and the one which was temporally the closest associated with the first observed SEP onset. 
For each flare the catalog records the time of the flare, its Carrington longitude and latitude, and the GOES class that is based on the X-ray observations. In cases of flares for which Solar Orbiter was viewing the backside of the Sun (i.e. flares not observed with GOES), an estimated GOES class is given based on the counts in the STIX 4-10 keV channel. This estimate is based upon a statistical correlation between the GOES 1-8~\AA\ flux and the STIX 4-10 keV channel for flares observed by both for all events from Jan 2021 - Nov 2023 \citep[see][ for further details]{stixdatacenter}. We note, however, that this temporal association does not exclude other potential flares in longitudinal sectors not observed with X-ray telescopes that could have contributed to the multi-S/C SEP observations. The user is therefore advised to take this possibility into account and if necessary to analyze the events in more detail themselves.
% However, a related catalog produced by the SERPENTINE project will provide a list with all flares that might have contributed to the observed SEP events.

%%%%%%%%%%%%%%%%%%%%%%%%%%%%%%%%%%%%%%%%%%%%%%%%%%%%%%%%%%%%%%%%%%%%%%%%
\subsection{Radio type-II-burst identification}\label{sec:typeII}

For each event in the SEP event catalog, the presence of metric and decametric type II radio bursts was investigated. For the identification of interplanetary type II bursts in the decametric range we visually inspected daily radio spectrograms\footnote{We used the plots provided at \url{https://parker.gsfc.nasa.gov/crocs.html}.} of Parker Solar Probe, STEREO~A, Solar Orbiter, and Wind and list which spacecraft did observe the type II. For the metric type II bursts we used event lists provided by SWPC. In rare cases, where a type II burst was not present in the SWPC lists, we identified the burst from the eCALLISTO event lists\footnote{\url{http://soleil.i4ds.ch/solarradio/data/BurstLists/2010-yyyy_Monstein/}} that are used for confirmation. These lists contain information on solar eruptions including the presence of radio burst types and their properties obtained from dynamic spectra from the ground-based (GB) RSTN telescope network. We extracted only type II radio bursts as these are signatures of shock-accelerated electron beams at the Sun \citep[e.g.,][]{ne85,Jebaraj2020,Kouloumvakos21,Jebaraj2021,Morosan19,Morosan23} and can be used to indicate the presence of a coronal shock, which is relevant for SEP events. 

Corresponding bursts were identified based on a temporal association with the flare start time. We only considered those type II bursts that occurred at most one hour from the flare start time. For both metric and decametric type II bursts we provide timing and frequency information. In case of the decametric type II burst, we use the spacecraft that observed the brightest type II burst. In case of the metric type II bursts the parameters were extracted from the SWPC event lists. Following the identification of type II bursts associated with SEP events, we also investigated the availability of radio imaging with the Low Frequency Array \citep[LOFAR;][]{vanHaarlem2013} and the Nan{\c c}ay Radioheliograph \citep[NRH;][]{Kerdraon1997}. The availability of radio imaging and the imaging instrument are provided in the column of the radio comments.

%%%%%%%%%%%%%%%%%%%%%%%%%%%%%%%%%%%%%%%%%%%%%%%%%%%%%%%%%%%%%%%%%%%%%%%%%
\subsection{Determination of SEP parameters}\label{sec:sep_parameters}
SEP onset times were determined with the Poisson-CUSUM method \cite[][]{Huttunen2005} using the SEP analysis tools of the SERPENTINE project \cite[][]{Palmroos2022}.
Figure~\ref{fig:multi-sc1} shows an example of two consecutive SEP events observed on 24 and 25 Feb 2023. The top plot shows the longitudinal configuration of the spacecraft fleet with respect to the associated flare (called 'reference longitude') for the first of the two SEP events. The bottom plot shows the intensity time series of the three energy / species combinations of the catalog: $\sim100$~keV electrons (top), $\sim1$-MeV electrons (middle), and $>25$~MeV protons (bottom). 
Onset and peak times are marked by solid and dotted lines, respectively, with the color coding according to the different spacecraft. We define the event's peak as the global intensity maximum, which can in some cases be strongly delayed with respect to the onset time. 
For each onset and peak time the catalog provides the time resolution that was used to determine these parameters. We aimed at using 1-minute averages for the onset determination if the onset is well-defined. For the peak time and peak intensity we used at least a time averaging of 10 minutes. However, if the onset determination required more time averaging, that is, more than 10 minutes, we used at least the same averaging as the one applied for the peak determination. 

If the used instrument provided different viewing directions (cf. Table~\ref{tab:instruments}), we used the viewing direction that observed the earliest onset time for the onset determination and that with the highest maximum intensity for the peak determination. The respective employed viewing directions are also provided in the catalog. 

In case of late intensity maxima, which could be associated with interplanetary shock acceleration, the 100-keV electron measurements can suffer from contamination by ions, which can dominate during later phases of SEP events. In those cases we excluded the 100-keV-electron peak times and intensities. This information is also marked in the respective comments column.

Although the first event in Figure~\ref{fig:multi-sc1} was observed by four observers (L1, STEREO~A, Solar Orbiter, and BepiColombo), it was not classified as a widespread event because none of the observers' magnetic footpoints at the Sun has a longitudinal separation with the flare greater than 80 degrees, which would prove a wide SEP spread. 

In case an event was not observed by a certain S/C, the catalog still provides the coordinates of the observer location. The value in the peak intensity column then corresponds to the background intensity observed by the respective S/C during the time of the event observed by other S/C, but no peak time is provided.\footnote{In case of Parker observations we do not provide the background intensity for electrons as those data were not made public yet at the time of writing.} This allows the user to estimate if the S/C was situated in an environment with increased SEP background levels. A spacecraft that did not observe an event also serves as an upper limit for the longitudinal extent of the SEP event. Figure~\ref{fig:multi-sc2}, for example, shows a narrow-spread event as constrained by Solar Orbiter, which did not detect the event.

For each of the three energy species combinations we also provide an inferred solar injection time based on a simple time-shift analysis \citep[e.g.,][]{Paassilta2018}. Therefore, the onset time is shifted back in time based on the expected travel time of the particles based on their speeds and an assumed path length along a nominal Parker spiral field line. To determine the length of the Parker spiral we use a one-hour average of the measured solar wind speed at the onset time of the event. If no solar wind speed measurement was available, a speed of 400 km/s is used. These measured solar wind speeds are provided in the catalog for each of the three particle types, as their onset times might be different.

%%%%%%%%%%%%%%%%%%%%%%%%%%%%%%%%%%%%%%%%%%%%%%%%%%%%%%%%%%%%%%%%%%%%%%%%
\subsection{Determination of electron to proton ratios}\label{sec:ratios}
The catalog contains the electron-to-proton intensity ratio of $\sim 1$-MeV electrons and 25--40~MeV protons. 
The instruments, energy channels, and corresponding mean energies used to determine the e/p ratios are summarized in Table~\ref{tab:e2p}. If not stated otherwise, the mean energies were calculated using a geometric mean of the energy range limits. 
For the protons we used the peak intensities that are directly provided in the catalog. The mean energies of the used energy channels provided by the different instruments are very similar at $\sim$32~MeV. For electrons we tried to match an energy of $\sim1$~MeV as best as possible. This required, however, a more complex utilization of instrument data products, which is described in the following. In case of STEREO/HET, we used channel 0, which has a geometric mean energy of 0.99~MeV. 
For SolO/EPD, the energy of 0.99~MeV is situated between the energy ranges covered by the EPT and HET instruments. We therefore used both telescopes and determined the $\sim$~1-MeV peak intensity using an interpolation between peak intensities determined for channel 1 of HET (1.053–-2.401~MeV) and the combined channel 31–33 for EPT (0.37--0.47~MeV) assuming a power law. In those cases where the event was not observed in the HET energy channel we extrapolate the EPT peak intensity using an average slope determined from 26 events, where this was possible. The average slope and its standard deviation are $-4.87\pm1.72$.

\begin{table*}
\caption{\label{tab:e2p} Energetic particle instruments, energy channels, and corresponding mean energies used to determine the e/p ratios.}
\centering
\begin{tabular}{lllll}
\hline\\
(1) & (2) & (3)& (4)& (5)\\[0.5mm]
                        & \multicolumn{2}{c}{25--40 MeV protons} & \multicolumn{2}{c}{1 MeV electrons}  \\
Spacecraft / Instrument & Energy channels (CH) & Mean energy $\langle E\rangle$ & Energy channel(s) & Mean energy $\langle E\rangle$\\[1mm]
 & & & used (CH) & of determined peak int. \\
 \hline \\
SOHO / ERNE / HED & CH 3--4:  & & &   \\
                  & 25 -- 40 MeV & 31.6 MeV & &  \\                
SOHO / EPHIN      &      &        & E5: 0.45 -- 0.5 MeV&    \multirow{2}{*}{$\left.\begin{array}{r} 
                \\
                \\
                \end{array}\right\rbrace 0.99~~MeV\tablefootmark{a}$} \\   
                  &      &        & E15: 0.7 -- 1.1 MeV &   \\
STEREO / HET      &  CH 5--8:     &   & CH 0:  \\
                  & 26.3 -- 40.5 MeV & 32.6 MeV & 0.7-1.4 MeV &  0.99 MeV \\
SolO / EPD / HET  &   CH 19--24:  & & CH 0-1: & \multirow{4}{*}{$\left.\begin{array}{r}
                \\
                \\
                \\
                \end{array}\right\rbrace 0.99~~MeV\tablefootmark{b}$} \\
                  & 25.09 -- 41.18 MeV & 32.1 MeV   & 0.45--2.4 MeV &\\
SolO / EPD / EPT  &       &        & CH 31--33:  & \\
                  &       &        & 0.37 -- 0.47 MeV & \\

Parker / EPI-Hi / HET & CH 8--9: & & CH 3--4: & \\
 & 26.91 -- 38.05 MeV & 32.0 MeV &0.71 -- 1.41 MeV & 1.00 MeV\\
 
BepiColombo / SIXS & CH 8--9: & & CH 5: &\\
                  & 25.1 -- 37.3 MeV\tablefootmark{c} & 35~MeV & 0.96 MeV\tablefootmark{c} & 0.96~MeV \\
  
\hline
\end{tabular}
\tablefoot{
\tablefoottext{a}{Correction and new data product used, see text. }
\tablefoottext{b}{Interpolation used.}
\tablefoottext{c}{Only effective energies available.}
}
\end{table*}

In case of SOHO/EPHIN we made use of a new data product. While the previously used level2 data product was relying on detector combinations with only limited energy resolution for electrons, the other extreme was the often used so-called Pulse Height Analysis (PHA) data, which provides detailed energy losses in each detector for only a statistical sample of measured particles. Here, we use neither of those but the onboard histogram data, which provides a good compromise between counting statistics and energy resolution. Utilizing the bow-tie method \citep[e.g.,][]{Raukunen2020} we were able to derive several energy channels based on this data set in the energy range between 0.15~MeV to about 1.1~MeV. In this work we make use of two of those new energy channels, E5 (0.45--0.5~MeV) and E15 (0.7--1.1~MeV).
%, which overcomes the current limitation due to instrument aging that resulted in a reduced number of energy channels in the level 2 data product. 
The effective mean energy of the E15 channel is 0.955~MeV. To match the desired 0.99~MeV even better, we interpolate between the two intensity values assuming a spectral slope of a power law just like for the Solar Orbiter measurements (see above). Because we require the intensity peaks in both energy channels to occur within two neighboring time steps based on the used time resolution, some events did not qualify for the interpolation method. In those cases we correct the peak intensity of E15 assuming a power law with a spectral slope that had been determined from the 17 events in which the interpolation method could be applied. The used average slope and its standard deviation are $-2.46\pm0.46$.

For BepiColombo we use the electron channel 5, which has a nominal mean energy of 0.96~MeV. We note that no exact energy ranges can be provided for BepiColombo/SIXS due to the complex response functions of the instrument. The mean energy provided in Table~\ref{tab:e2p} has been determined using a bow-tie analysis with simulated instrument response function \citep{Huovelin2020}. 

%We note furthermore that we were not able to determine e/p ratios for Parker at this time due to the electron data available only in count rates instead of intensities.

\section{Results} \label{sec:results}
\begin{figure*}[]  
   \centering
        \includegraphics[width=0.95\textwidth]{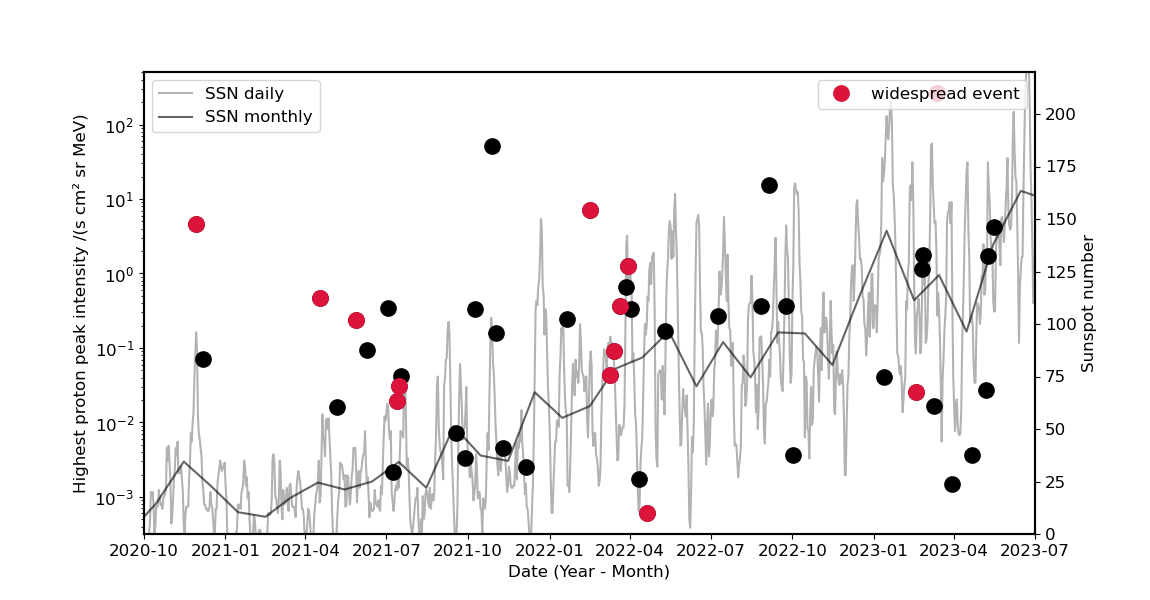} \\
        \includegraphics[width=0.95\textwidth]{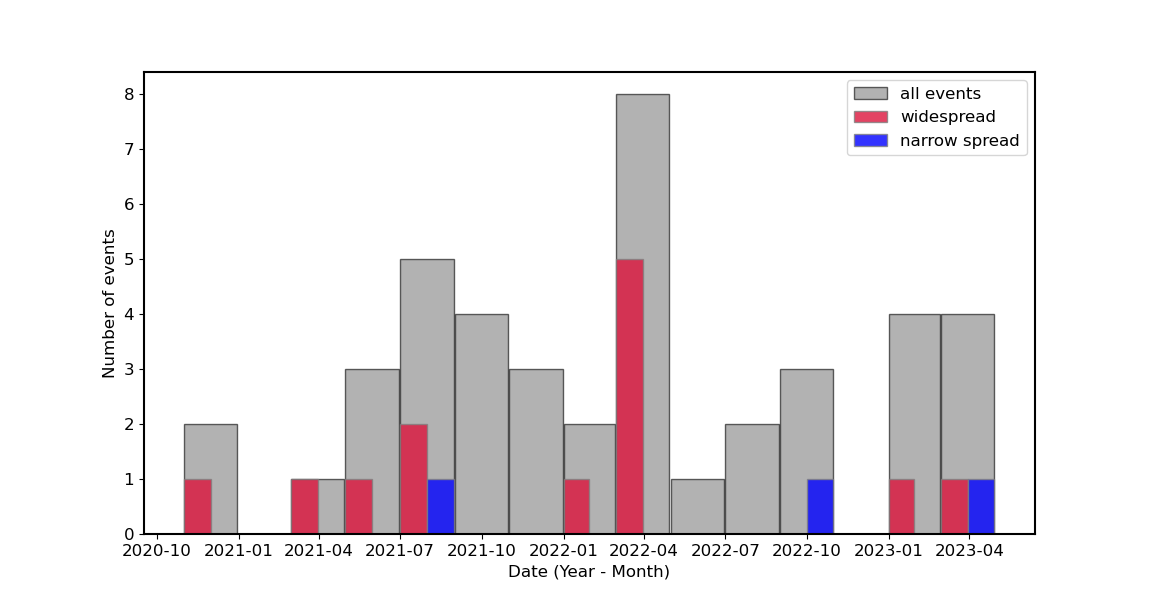}  
        \caption{Top: Highest $>25$~MeV-proton peak intensity observed (black dots) in each event as a function of time with widespread events marked in red. Daily and monthly sunspot numbers are shown by gray curves (right axis) provided by WDC-SILSO, Royal Observatory of Belgium, Brussels. Bottom: Histogram of event occurrences with all events in gray, widespread events in red, and narrow-spread events in blue.} 
        \label{fig:hist_event_dates}
\end{figure*}

The current version of the catalog contains 45 events covering the period from Nov 2020 to May 2023. The histogram shown in Figure~\ref{fig:hist_event_dates} (bottom) shows the number of events over time and indicates when widespread and narrow-spread events were observed. The top plot of Figure~\ref{fig:hist_event_dates} shows the highest proton peak intensity observed in each event as a function of time, marking all widespread events in red. We neither observe a clear correspondence between peak intensities and widespread events nor between the highest peak intensities and sunspot numbers, which are also shown in the plot. 

\begin{figure}[]  
    \centering
        \includegraphics[width=0.48\textwidth]{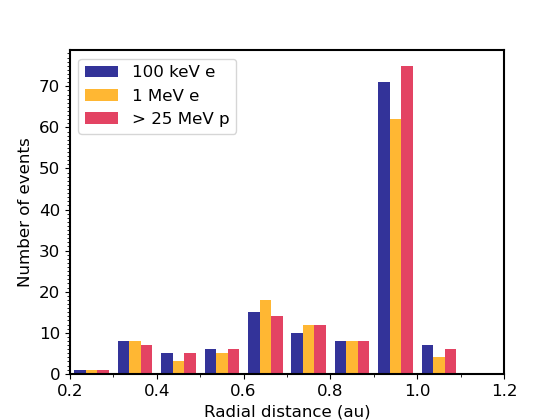} 
        \caption{Histogram of spacecraft distances during event observations for the three different energy-species combinations.} 
        \label{fig:hist_dist}
\end{figure}

Counting all single spacecraft SEP event observations of the current version of the catalog yields 142 events in total. Counting each energy/species combination separately, we get 414 events for which the catalog provides key parameters. Figure~\ref{fig:hist_dist} shows the radial distances at which the S/C observed these events. The slight differences in the event numbers based on the different energy / species combinations can be caused by a few events being observed only at lower energies but not extending to higher energies, or by data gaps. 
With Parker, Solar Orbiter, and BepiColombo, events were also observed at smaller heliocentric distances, with 20 (45) single-S/C events at radial distances below 0.6~AU (0.8~AU).

\begin{figure}[]  
    \centering
        \includegraphics[width=0.5\textwidth]{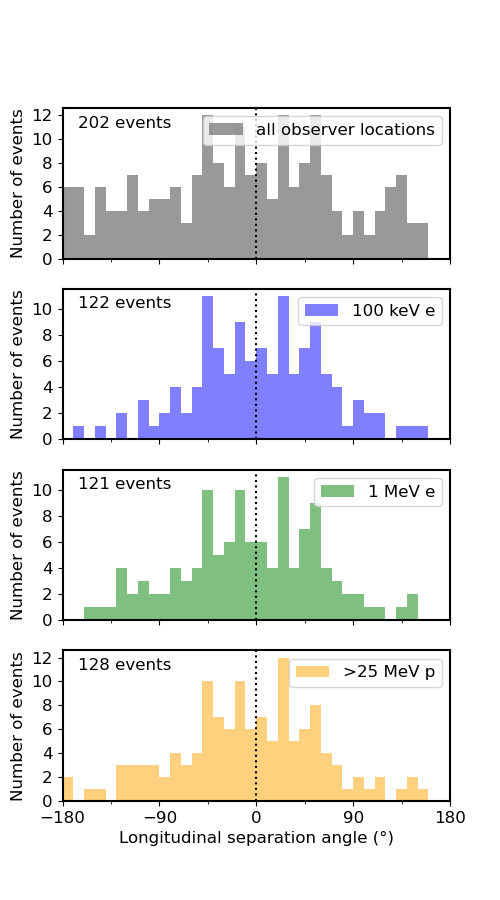} 
        \caption{Distribution of longitudinal separation angles between observers' magnetic footpoints and associated flare location. Positive (negative) separation angles denote a magnetic footpoint connecting western (eastern) of the flare longitude.
        Top: all events, including separation angles of observers not observing an event. Below: Longitudinal separation angles of observers who observe a $\sim100$-keV electron event (blue), a $\sim1$-MeV electron event (green), and a $>25$-MeV proton event (yellow).} 
        \label{fig:hist_longsep}
\end{figure}

Figure~\ref{fig:hist_longsep} (top) shows a histogram of the longitudinal separation angles between spacecraft magnetic footpoints at the Sun and flare location of all spacecraft for all events in the catalog, even if no associated SEP event was observed. The three panels below show the longitudinal separation angles only for those observers who detected a $\sim100$-keV electron, $\sim1$-MeV electron, or $>25$-MeV proton event. These lower three panels show very similar Gaussian-like distributions indicating that an SEP event is only rarely observed at longitudinal separation angles larger than 100 degrees. Because only events, for which the flare location was known, could be included in Figure~\ref{fig:hist_longsep}, the total number of events (stated in the figure legend) is slightly lower here. 

\begin{figure*}[]  
    \centering
        \includegraphics[width=0.3\textwidth, trim={0, 0, 80, 0}, clip]{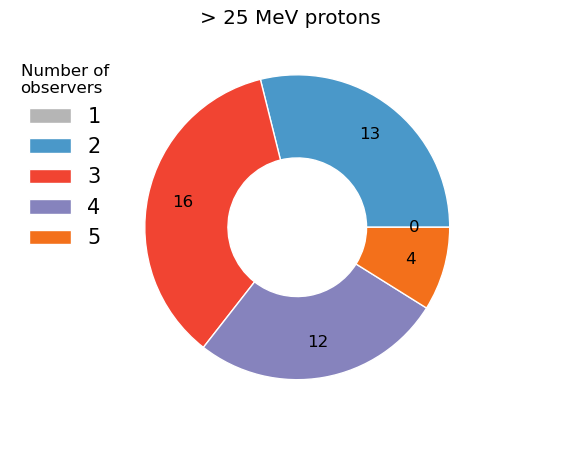} 
        \includegraphics[width=0.3\textwidth, trim={0, 0, 80, 0}, clip]{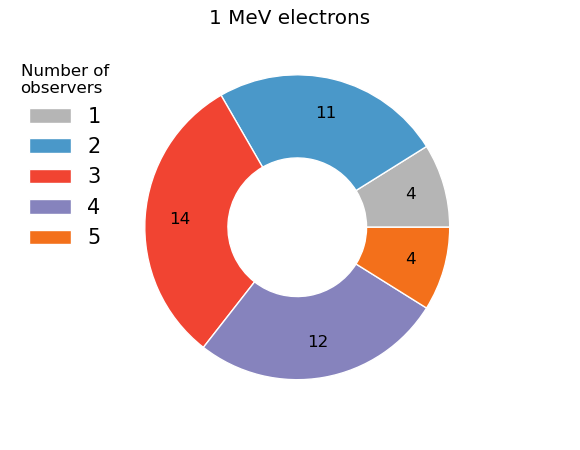} 
         \includegraphics[width=0.3\textwidth, trim={0, 0, 80, 0}, clip]{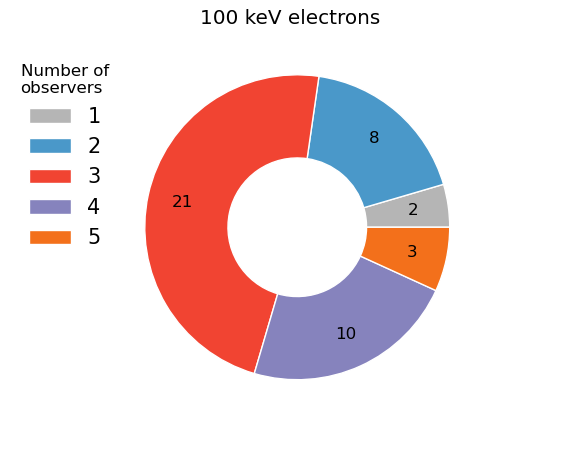} 
        \caption{Number of observers per event for $>25$~MeV protons (left), $\sim1$-MeV electrons (middle), and $\sim100$~keV electrons (right). } 
        \label{fig:donuts}
\end{figure*}

Figure~\ref{fig:donuts} shows how many events were observed by how many S/C. The majority of events were observed by three S/C. In case of the $\sim$100-keV electrons, even more 3-S/C events and less 2-S/C events were observed than for the higher energy particles.

\begin{figure*}[]  
    \centering
        \includegraphics[width=0.3\textwidth, trim={0, 0, 80, 0}, clip]{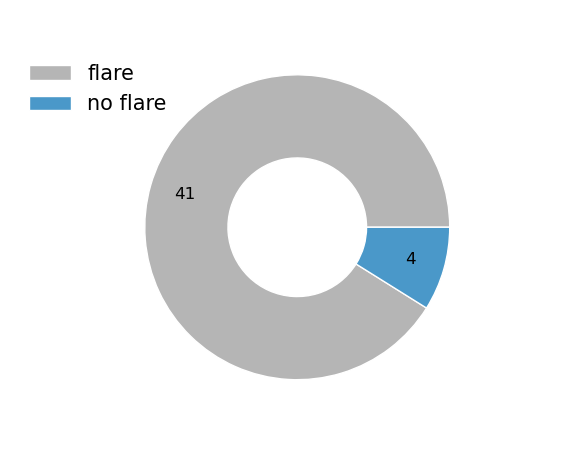} 
        \includegraphics[width=0.3\textwidth, trim={0, 0, 80, 0}, clip]{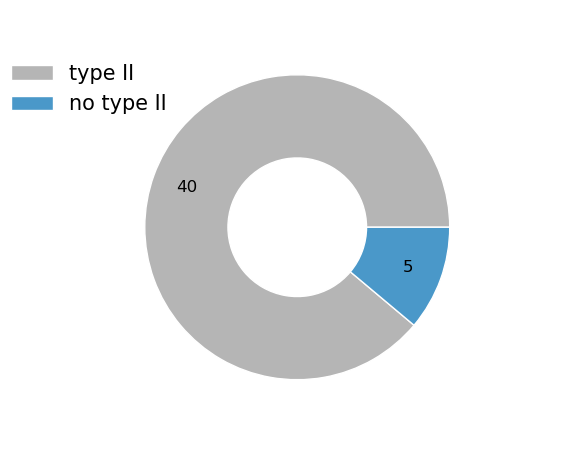} 
        \includegraphics[width=0.3\textwidth, trim={0, 0, 80, 0}, clip]{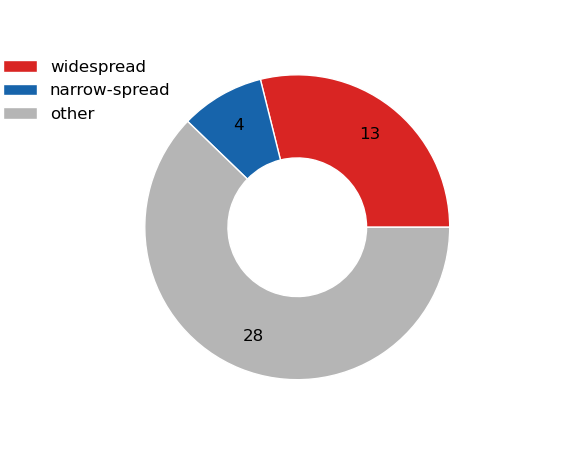}
        \caption{Number of events for which an associated flare (left) or an associated type-II radio burst (center) was identified. The right-hand plot depicts the fraction of events, which showed a widespread or narrow-spread SEP event.} 
        \label{fig:donuts2}
\end{figure*}

Figure~\ref{fig:donuts2} shows the fraction of events for which we found an associated flare (left) or type-II radio burst (center). The four missing flares are most likely caused by missing X-ray coverage at these specific flare locations. Most events in the catalog were associated with a shock, as marked by the larger number of 40 type-II radio bursts out of the total 45 events. The right-hand plot of Figure~\ref{fig:donuts2} shows the number of events for which we could infer a wide or narrow SEP spread. Because our ability to characterize the absolute spread strongly depends on the individual observer positions, most of the events (shown in gray) could not be characterized in this way.

%\section{Related catalogs of the SERPENTINE project} %\label{sec:other_catalogs}
%

\section{Summary} \label{sec:summary}
We present a new multi-S/C catalog of SEP events detected during solar cycle 25 that reached proton energies of 25~MeV as observed by at least two different spacecraft in the inner heliosphere. The catalog contains key SEP parameters (see Sect.~\ref{sec:catalog}) for $\sim$25--40 MeV protons as well as $\sim$100-keV and $\sim1$-MeV electrons utilizing five different observer locations provided by six space missions, that is, Solar Orbiter, Parker, STEREO~A, BepiColombo, SOHO, and Wind, with the latter two both being situated close to the Lagrangian point L1.

Aiming at a most accurate determination of SEP parameters such as onset and peak times, we individually chose the most appropriate time averaging and make use of different viewing directions of the instruments, if those are available, and used the viewing direction which was best aligned with the magnetic field during each respective SEP event.

For each event, the location of all spacecraft is provided, regardless of whether one detected the SEP event or not. We also provide the background intensity level in case an observer does not detect the event. 

Although each of the SEP events of the catalog is a multi-spacecraft event, not all of these are observed over a wide longitudinal range. The catalog therefore contains a variety of science cases, for example widespread SEP events, narrow-spread events, or events with some observers magnetically or radially aligned. 

The catalog is provided online as a .csv-file via a citable Zenodo entry \citep{Dresing2024zenodo} and at the SERPENTINE data center. The latter provides in addition a more sophisticated web-interface, allowing to apply various filtering options and to explore the links with other event catalogs provided by the SERPENTINE project, which have a supporting role of the SEP event catalog. 
These are in-situ shock catalogs as well as a catalog of CMEs, which were potentially associated with the SEP events. 
Also available at the SERPENTINE data center are two historic-data catalogs based on observations of the Helios spacecraft; a SEP event catalog and an in-situ shock catalog.

\begin{acknowledgements}
We acknowledge funding by the European Union’s Horizon 2020 research and innovation program under grant agreement No.\ 101004159 (SERPENTINE). ND is grateful for support by the Academy of Finland (SHOCKSEE, grant No.\ 346902). DEM acknowledges the Academy of Finland project 'SolShocks' (grant number 354409).     
The computer resources of the Finnish IT Center for Science (CSC) and the FGCI project (Finland) are acknowledged. 
Some of the Parker data analysis is supported by NASA’s Parker Solar Probe Mission, contract NNN06AA01C. Parker Solar Probe was designed, built, and is now operated by the Johns Hopkins Applied Physics Laboratory as part of NASA’s Living with a Star (LWS) program.  The ISOIS data are available to the community at https:// spacephysics.princeton.edu/missions-instruments/isois; data are also available via the NASA Space Physics Data Facility (https://spdf. gsfc.nasa.gov/). 
BH, PK and SJ thank the German Federal Ministry for Economic Affairs and Energy and the German Space Agency (Deutsches Zentrum für Luft- und Raumfahrt, e.V., (DLR)) for their support of Solar Orbiter EPT and HET, STEREO SEPT and SOHO EPHIN under grants number 50OT2002 and 50OC2102. 
The UAH team acknowledges the financial support by the Spanish Ministerio de Ciencia, Innovación y Universidades project PID2019-104863RBI00
/AEI/10.13039/501100011033.
The AIP team  was supported by the German Space Agency (DLR) under grant numbers \mbox{50 OT 1904} and \mbox{50 OT 2304}.
BH acknowledges the support of German research foundation under grant number GI 1352/1-1.
We also acknowledge SolarMonitor.org, the online catalog of flares observed by the Hinode satellite, and SpaceWeather prediction Center for providing information about solar flares included in the present catalog.
EA acknowledges support from the Academy of Finland/Research Council of Finland (Academy Research Fellow grant number 355659-Project SOFTCAT).
\end{acknowledgements}

\bibliography{references}{}
\bibliographystyle{aa}
%
%
%-------------------------------------------------------------
%                   For appendices and landscape, large table:
%                    in the preamble, use: \usepackage{lscape}
%-------------------------------------------------------------

% \begin{appendix} %First appendix

% \end{appendix}

\end{document}